\documentclass[onecolumn,superscriptaddress,showpacs,nofootinbib,notitlepage]{aastex631}
\usepackage{graphicx,subfigure,amsmath,amsfonts,amssymb,multirow,mathrsfs,ulem,threeparttable}
\usepackage{hyperref}
\usepackage{epstopdf}
\begin{document}
\title{The late afterglow of GW170817/GRB170817A: a large viewing angle and the shift of the Hubble constant to a value more consistent with the local measurements}

\correspondingauthor{Yi-Zhong Fan}
\email{yzfan@pmo.ac.cn}
\correspondingauthor{Zhi-Ping Jin}
\email{jin@pmo.ac.cn}
\author[0000-0003-1215-6443]{Yi-Ying Wang}
\affiliation{Key Laboratory of Dark Matter and Space Astronomy, Purple Mountain Observatory, Chinese Academy of Sciences, Nanjing 210033, People's Republic of China}
\affiliation{School of Astronomy and Space Science, University of Science and Technology of China, Hefei, Anhui 230026, People's Republic of China}
\author[0000-0001-9120-7733]{Shao-Peng Tang}
\affiliation{Key Laboratory of Dark Matter and Space Astronomy, Purple Mountain Observatory, Chinese Academy of Sciences, Nanjing 210033, People's Republic of China}
\author[0000-0003-4977-9724]{Zhi-Ping Jin}
\affiliation{Key Laboratory of Dark Matter and Space Astronomy, Purple Mountain Observatory, Chinese Academy of Sciences, Nanjing 210033, People's Republic of China}
\affiliation{School of Astronomy and Space Science, University of Science and Technology of China, Hefei, Anhui 230026, People's Republic of China}
\author[0000-0002-8966-6911]{Yi-Zhong Fan}
\affiliation{Key Laboratory of Dark Matter and Space Astronomy, Purple Mountain Observatory, Chinese Academy of Sciences, Nanjing 210033, People's Republic of China}
\affiliation{School of Astronomy and Space Science, University of Science and Technology of China, Hefei, Anhui 230026, People's Republic of China}

\newcommand{\ud}{\mathrm{d}}
\begin{abstract}
The multi-messenger data of neutron star merger events are promising for constraining the Hubble constant. So far, GW170817 is still the unique gravitational wave event with multi-wavelength electromagnetic counterparts. In particular, its radio and X-ray emission have been measured in the past $\sim 3-5$ years. In this work, we fit the long-lasting X-ray, optical, and radio afterglow light curves of GW170817/GRB 170817A, including the forward shock radiation from both the decelerating relativistic GRB outflow and the sub-relativistic kilonova outflow (though whether the second component contributes significantly is still uncertain), and find out a relatively large viewing angle ($\sim 0.5\, \rm rad$). Such a viewing angle has been taken as a prior in the gravitational wave data analysis, and the degeneracy between the viewing angle and the luminosity distance is broken. Finally, we have a Hubble constant  $H_0=72.57^{+4.09}_{-4.17}\, \rm km\, s^{-1}\, Mpc^{-1}$, which is more consistent with that obtained by other local measurements. If rather similar values are inferred from multi-messenger data of future neutron star merger events, it will provide critical support to the existence of the Hubble tension.
\end{abstract}

\section{Introduction}\label{sec:1}
The Hubble constant ($H_0$) is a fundamental parameter of cosmology. However, its measurements obtained with different methods are clustered into two groups \citep{2019NatAs...3..891V}: one is represented by the cosmic microwave background (CMB) measurement from the Planck Collaboration for the early universe ($67.66 \pm 0.42 {\rm ~km\,s^{-1}\,Mpc^{-1}}$; \citealt{2020A&A...641A...6P}) and the other is represented by the type Ia supernova and Cepheids measurements from the SH0ES (Supernova, $H_0$, for the Equation of state of Dark Energy) team in the local universe ($73.30 \pm 1.04 {\rm ~km\,s^{-1}\,Mpc^{-1}}$; \citealt{2021arXiv211204510R}). So far, it is still unclear whether such a severe tension is due to the presence of new physics (i.e., the modification of the standard cosmology model) or alternatively some unknown systematic bias introduced in the local measurements \citep{2021ApJ...912..150D,2022Galax..10...24D}. Independent precise local $H_0$ measurements without using the distance ladders are thus necessary to check the second possibility. The gravitational wave (GW) events are expected to play an important role in such an aspect since GW, serving as the standard siren, can be used to measure $H_0$ \citep{1986Natur.323..310S}. This is particularly the case for the neutron star merger events with detected multi-wavelength electromagnetic counterparts. For such events, the redshifts can be reliably measured, and the inclination angles (i.e., the viewing angle) of the mergers may be robustly inferred. Therefore, the degeneracy between the luminosity distance and the inclination angle can be effectively broken. The accurate redshift and luminosity distance thus lead to a direct measurement of $H_0$. 

The multi-messenger data of GW170817 \citep{2017ApJ...848L..12A, 2019PhRvX...9a1001A} has been extensively used to estimate the Hubble constant. \citet{2017Natur.551...85A} took the strain data and the redshift measurement of GW170817 to measure $H_0$ which was determined to be ${70}^{+12.0}_{-8.0} ~\rm km\,s^{-1}\,Mpc^{-1}$. Later, the inclusion of the electromagnetic radiation information yields a more accurate measurement of $H_0$, as reported in the literature. For instance, \citet{2019NatAs...3..940H} took both the superluminal motion and the multi-wavelength radiation of the relativistic jet into account and yielded a viewing angle of $0.29^{+0.03}_{-0.02}$ rad for the Power-law jet model, with which a $H_0=68.1^{+4.5}_{-4.3}\, \rm km\,s^{-1}\, Mpc^{-1}$ is reported. \citet{2021ApJ...908..200W} found that $H_0=69.48^{+4.3}_{-4.2}\, \rm km\,s^{-1}\,Mpc^{-1}$ when further including the direct measurement of the luminosity distance. There are also some constraints reported from other literature, e.g.,  $H_0=64.8^{+7.3}_{-7.2}\, \rm km\,s^{-1}\,Mpc^{-1}$ \citep{2020MNRAS.492.3803H}, $H_0=66.2^{+4.4}_{-4.2}\, \rm km\,s^{-1}\,Mpc^{-1}$ \citep{2020Sci...370.1450D}, and $H_0=68.3^{+4.6}_{-4.5}\, \rm km\,s^{-1}\,Mpc^{-1}$ \citep{2021A&A...646A..65M}. Though the uncertainties are relatively large, their median values are close to those found in the CMB, baryon acoustic oscillations, and Big Bang nucleosynthesis experiments \citep{2018MNRAS.480.3879A}.

However, in the previous afterglow-involved $H_0$ measurements, only the first-year afterglow data of GRB 170817A have been included in the modeling (for instance, \citet{2021ApJ...908..200W} just fitted the afterglow data collected in the first 200 days to infer the viewing angle of the GRB ejecta due to the lack of reliable treatment on the sideways expansion in the code). Currently, the afterglow observations of GRB 170817A have been accumulated to almost 4.8 years \citep{2022GCN.32065....1O}. These late time afterglow data are expected to constrain the physical parameters tightly. Interestingly, it is found that the viewing angle $\theta_{\rm v}$ of the GRB ejecta (which is widely assumed to be the same as the inclination angle $\iota$ of the merger event) gets increased if the late time data have been included in the fit \citep{2019ApJ...886L..17H, 2021MNRAS.502.1843N, 2022ApJ...927L..17H}. Therefore, it is necessary to fit all the available afterglow data to yield a more reliable $\theta_{\rm v}$ (i.e., $\iota$). Besides, the superluminal motion of a relativistic jet of GW170817 \citep{2018Natur.561..355M, 2019Sci...363..968G, 2019NatAs...3..940H} that provides an extra constraint on $\theta_{\rm v}$ will also be incorporated in this work. By performing the Bayesian analysis on both the multi-wavelength light curves and GW data, we find that the Hubble constant is $H_0=72.57^{+4.09}_{-4.17}\, \rm km\, s^{-1}\, Mpc^{-1}$, which is more consistent with that obtained by other local measurements.

\section{The Afterglow Data}\label{sec:2}
Recently, the synchrotron afterglow emission of GRB 170817A in X-ray wavelength after 1674 days since the merger of GW170817 was detected \citep{2022GCN.32065....1O}. Therefore, we incorporate this new observation with all of the available data in radio, optical, and X-ray bands into the afterglow modeling (which will be described in Section.~{\ref{sec:3}}). In the radio band, the data cover the early and late duration from 16 to 1243 days after the BNS merger which are obtained by the Karl G. Jansky Very Large Array \citep[VLA;][]{2017Sci...358.1579H, 2018ApJ...863L..18A, 2018ApJ...856L..18M, 2018Natur.561..355M}, the Australia Telescope Compact Array \citep[ATCA;][]{2017Sci...358.1579H, 2018ApJ...858L..15D, 2018ApJ...868L..11M, 2018Natur.554..207M, 2021ApJ...922..154M}, the Giant Metrewave Radio Telescope \citep[uGMRT;][]{2018ApJ...867...57R, 2018ApJ...868L..11M}, the enhanced Multi Element Remotely Linked Interferometer Network \citep[eMERLIN;][]{2021ApJ...922..154M}, the Very Long Baseline Array \citep[VLBA;][]{2019Sci...363..968G}, and the MeerKAT telescope \citep{2018ApJ...868L..11M, 2021ApJ...922..154M}. In the optical band, the data distributed around the light-curve peak from 109 to 362 days post-merger are obtained by the Hubble Space Telescope \citep[HST;][]{2018NatAs...2..751L, 2019ApJ...883L...1F, 2019ApJ...870L..15L}. The observations in the X-ray band from the early (9 days) to the very late (1674 days) time after the merger are obtained by the XMM-Newton \citep{2018A&A...613L...1D, 2019MNRAS.483.1912P} and Chandra \citep{2017Natur.551...71T, 2018MNRAS.478L..18T, 2019ApJ...886L..17H, 2020MNRAS.498.5643T, 2022GCN.32065....1O, 2022MNRAS.510.1902T}.

\section{Method}\label{sec:3}
As the only ``standard siren" so far, the source of GW170817 is confirmed to be located in NGC 4993 \citep{2017ApJ...848L..12A}. In the local universe, we have Hubble's law,
\begin{equation}\label{eq:hubble_law}
	v_{\rm H}=v_{\rm r}-v_{\rm p}=H_0d_{\rm L},
\end{equation}
where $v_{\rm H}$ is the local Hubble flow velocity of the galaxy, $v_{\rm r}$ is the recession velocity of the galaxy relative to the CMB frame, and $v_{\rm p}$ is the peculiar velocity of the galaxy. Thus, the posterior probability of $H_0$ is
\begin{equation}\label{eq:standard_llh}
    p(H_0|x_{\rm GW},\langle v_{\rm H}\rangle)\propto p(H_0)\int \ud d_{\rm L} \ud v_{\rm H} p(v_{\rm H})p(\langle v_{\rm H}\rangle|v_{\rm H})p(d_{\rm L}|x_{\rm GW}),
\end{equation}
where $p(d_{\rm L}|x_{\rm GW})$ is the posterior distribution of $d_{\rm L}$ given by the Bayesian analysis on $x_{\rm GW}$, and $p(\langle v_{\rm H}\rangle|v_{\rm H})$ is the likelihood of Hubble flow velocity measurement. Here, we use the result from \citet{2021A&A...646A..65M}, i.e., the velocity of the Hubble flow is $v_{\rm H}=2954\pm148.6 \rm~km \, s^{-1}$ following a Gaussian distribution. 

To obtain a high precision luminosity distance from the GW data analysis, we need to break the degeneracy between $\iota$ and $d_{\rm L}$. Therefore, how to (independently) acquire a reliable measurement of $\iota$ is crucial. Assuming that the viewing angle $\theta_{\rm v}$ in GRB afterglow is equal to the inclination angle $\iota$, an acknowledged method is to infer the angle by fitting multi-light curves with afterglow models. In this work, we adopt two approaches (i.e., the {\tt Afterglowpy} and {\tt JetFit}) developed by \citet{2020ApJ...896..166R} and \citet{2018ApJ...869...55W} to constrain $\theta_{\rm v}$, respectively. For {\tt Afterglowpy}, multi-numerical/analytic structured jet models are implemented for calculating GRB afterglow light curves and spectra. While for {\tt JetFit}, since there are $\sim$2,000,000 synchrotron spectra computed from the hydrodynamic simulations, the full parameter space for fitting the light curves can be well explored by Markov chain Monte Carlo analysis \citep{2018ApJ...869...55W}. In this work, we take the model from \citet{2018ApJ...869...55W} as the fiducial one because the {\tt Afterglowpy} code makes approximation\footnote{To simplify the calculation, {\tt Afterglowpy} assumes a sideways expansion rate of the local sound speed. However, the numerical hydrodynamical simulations \citep{2003ApJ...591.1075K} revealed that the sideways expansion rate is usually lower than the speed of sound, which predicts shallower-decaying light curves at late times.} on the sideways expansion; moreover, the {\tt JetFit} gives stronger Bayes evidence, as we will show below. Except for GRB afterglow, kilonova afterglow might become a dominant component \citep{2022ApJ...927L..17H} at late times. Driven by the kilonova blast wave, the kilonova afterglow is caused by a shock through the external medium. Therefore, it can be approximated as a spherical cocoon model in the sub-relativistic regime. Lately, \citet{2022MNRAS.516.4949S} exploited an open source package {\tt Redback} for fitting electromagnetic transient. They approximated the kilonova afterglow as a spherical cocoon afterglow extended in {\tt afterglowpy} but with some constraints. As shown in Table~\ref{tb:Ag_par}, we adopt the setting of the prior of the kilonova afterglow in {\tt Redback} but import {\tt afterglowpy} directly. More complicated models of the kilonova afterglow and further discussions can be found in \citet{2019MNRAS.487.3914K} and \citet{2021MNRAS.506.5908N}. Therefore, such a component, calculated with the {\tt Afterglowpy} code, is also incorporated in our numerical fits. 

We update the analysis with multi-band light-curve data of GRB 170817A, including the data from radio and optical bands to X-ray bands, from 9.2 to 1674 days. Following \citet{2018ApJ...869...55W}, the {\tt JetFit} model in our work has eight free parameters. Their prior distributions are summarized in Table~\ref{tb:Ag_par}\footnote{The same as those in the Table~2 of \citet{2018ApJ...869...55W}, except that the fraction of electrons accelerated by the shock is fixed to $1$ and the luminosity distance $d_{\rm L}$ follows a Gaussian distribution with $\mu=40.7$ and $\sigma=2.36$ Mpc \citep{2021A&A...646A..65M}}. For other structured jet models in {\tt Afterglowpy}, the priors are similar to those in Table~\ref{tb:Ag_par}.
Additionally, the superluminal motion of the jet observed with Very Long Baseline Interferometry (VLBI) gives a constraint of $0.25<\theta_{\rm v} \bigl(\frac{d_{\rm L}}{41 \rm Mpc} \bigr)<0.5$ rad \citep{2018Natur.561..355M}. The likelihood of fitting the observation data (assuming the measurement error follows Gaussian distribution) with the afterglow model in the Bayesian statistical framework can be written as
\begin{equation}
   {\rm Likelihood}=\prod^{N}_{i} \frac{1}{\sqrt{2\pi}\sigma_i} {\rm exp} \biggl[ -\frac{1}{2} \biggl(\frac{f(x_i)-y_i}{\sigma_i} \biggr)^2 \biggr],
\end{equation}
where $(x_i,y_i)$ and $\sigma_i$ are the observed light-curve data and their uncertainties, respectively. $f(x_i)$ is the value predicted by the afterglow model at $x_i$. {Balancing the accuracy and the efficiency, the Bayesian analysis of these afterglow parameters adopt {\tt Pymultinest} as the sampler.}

After obtaining the posterior distribution of inclination angle through the afterglow light-curve fitting, we can take the posterior distribution as a prior and input it into the Bayesian analysis of GW data. The priors of other GW parameters are shown in Table~\ref{tb:GW_par}, where the prior of $d_{\rm L}$ follows the distribution obtained by \citet{2021A&A...646A..65M}. We assume that the BNS has aligned spins, and the precession effects can be neglected. As for the waveform template, we use the model of {\tt IMRPhenomD\_NRTidal} \citep{PhysRevD.99.024029}. The calibration uncertainties may slightly impact Hubble constant measurements \citep{2022arXiv220403614H}. While for GW170817, we find that the difference between $H_0$ results obtained with and without considering calibration is negligible since the uncertainty of peculiar velocity still dominates the uncertainty of $H_0$. We calculate the marginalized posterior distribution of luminosity distance using the GW parameter inference code {\tt Bilby} \citep{2019ApJS..241...27A} and adopt {\tt dynesty} \citep{2020MNRAS.493.3132S} as the nest sampler. Given a uniform prior distribution from 20 to 140 $\rm km\, s^{-1}\, Mpc^{-1}$, the Hubble constant then can be well estimated using Equation~(\ref{eq:standard_llh}) based on the tight constraint of the luminosity distance.

\begin{table*}[ht!]
\begin{ruledtabular}
\centering
\caption{Prior distributions of the parameters for GRB 170817A and kilonova afterglow}
\label{tb:Ag_par}
\begin{tabular}{lcc}
Names                   &Parameters  &Priors of parameter inference    \\ \hline            
Explosion energy                     &$\log_{10} E_{0,50}$\textsuperscript{a}    &Uniform(-6, 7)\\
Circumburst density\textsuperscript{b}                  &$\log_{10} n_{0,0}$\textsuperscript{a}     &Uniform(-6, 0)\\
Asymptotic Lorentz factor            &$\eta_0$                 &Uniform(2, 10)\\
Boost Lorentz factor                 &$\gamma_B$               &Uniform(1, 12)\\
Spectral index\textsuperscript{b}                         &$p$                      &Uniform(2, 2.5)\\	
Electron energy fraction\textsuperscript{b}               &$\log_{10} \epsilon_e$     &Uniform(-6, 0)\\
Magnetic energy fraction\textsuperscript{b}               &$\log_{10} \epsilon_B$     &Uniform(-6, 0) \\
Viewing angle\textsuperscript{b}                      &$\theta_{\rm v}/\rm rad$       &Sine(0, $\pi$)   \\
Isotropic-equivalent energy\textsuperscript{b}        &$\log_{10} E_{\rm iso}/\rm erg$    &Uniform(-44, 57)\\
Half opening angle\textsuperscript{b}                 &$\theta_c$                 &Uniform(0, $\pi/2$)\\
Outer truncation angle\textsuperscript{b}                   &$\theta_w$               &Uniform(0, $\pi/2$)\\
Luminosity distance\textsuperscript{b}                    &$d_{\rm L}/\rm Mpc$                      &Gaussian($\mu=40.7, \sigma=2.36$) \\ \hline
Maximum 4-velocity of outflow         &$U_{\rm Max}$  &Uniform(0.15, 0.7)\\
Minimum 4-velocity of outflow         &$U_{\rm Min}$  &Uniform(0.1, 0.15)\\
Normalization of outflow's energy distribution  &$E_{\rm r}/\rm erg$  &Uniform(45,50)\\
Power-law index of outflow's injection          &$k$  &Uniform(0.5,4)\\
Mass of material at $U_{\rm Max}$     &$\log_{10}(\rm Eject \ mass)/\rm M_{\odot}$  &Uniform(45,50)\\
Spectral index                        &$p$       &Uniform(2.0, 2.5)\\
Electron energy fraction              &$\log_{10} \epsilon_e$     &Uniform(-5, 0)\\
Magnetic energy fraction               &$\log_{10} \epsilon_B$     &Uniform(-5, 0) \\
Fraction of electrons that get accelerated       &$\xi_{N}$        &Uniform(0,1)\\
Initial lorentz factor                &$\Gamma_0$     &Uniform(1,0)

\end{tabular}
\begin{tablenotes}
  \item[a] \textsuperscript{a} Note that, $E_{0,50} \equiv E_0/10^{50} \, \rm erg$ and $n_{0,0} \equiv n_0/1 \, \rm proton \, cm^{-3}$.
  \item[b] \textsuperscript{b} These parameters are used in Gaussian structured jet in {\tt Afterglowpy}.
\end{tablenotes}
\end{ruledtabular}
\end{table*}

\section{Result}\label{sec:4}
\begin{figure}[ht!]
	\centering	
    \subfigure[]{
    \includegraphics[width=0.46\textwidth]{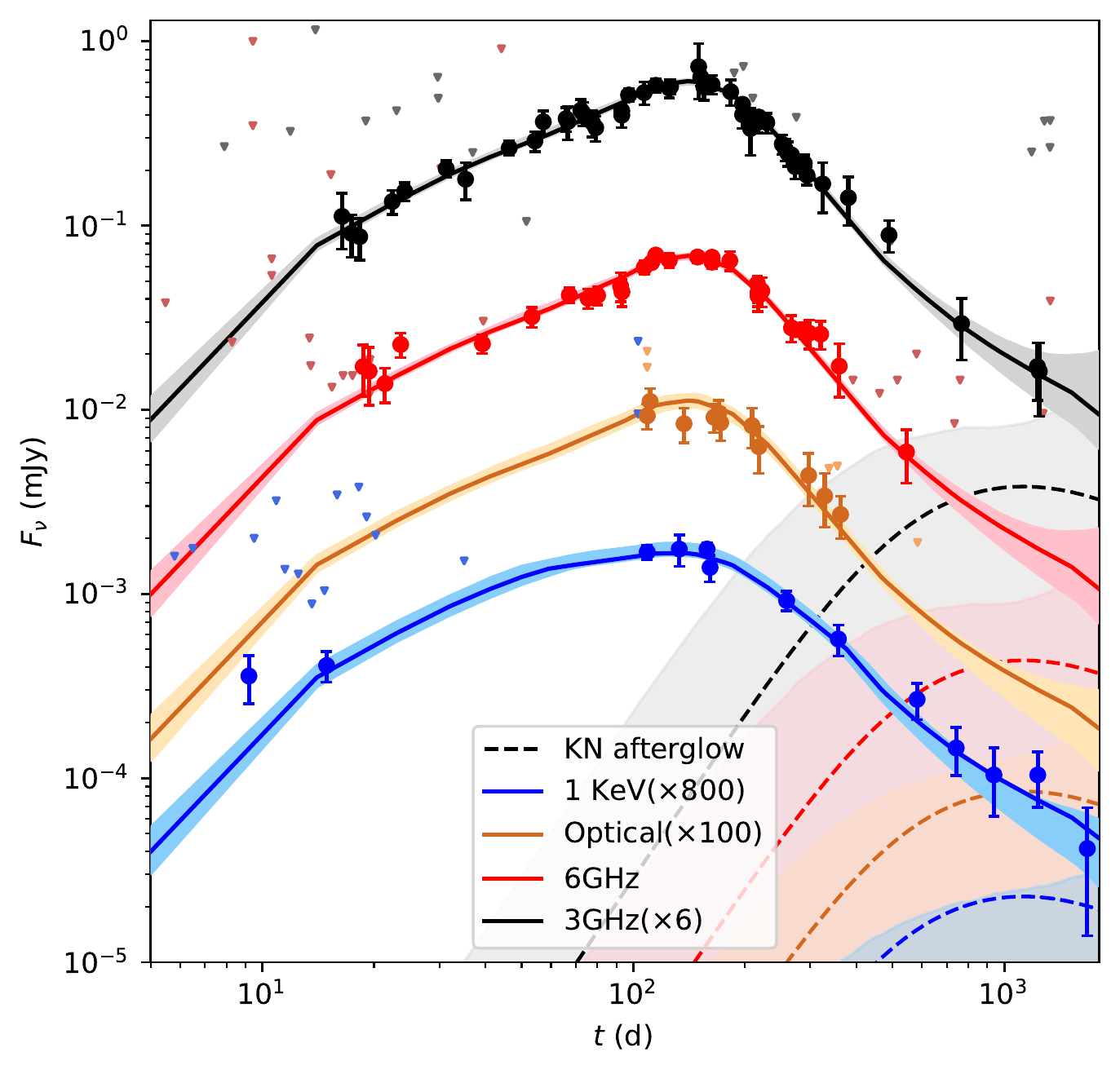}
    \label{subfig:1a}}
    \subfigure[]{
    \includegraphics[width=0.46\textwidth]{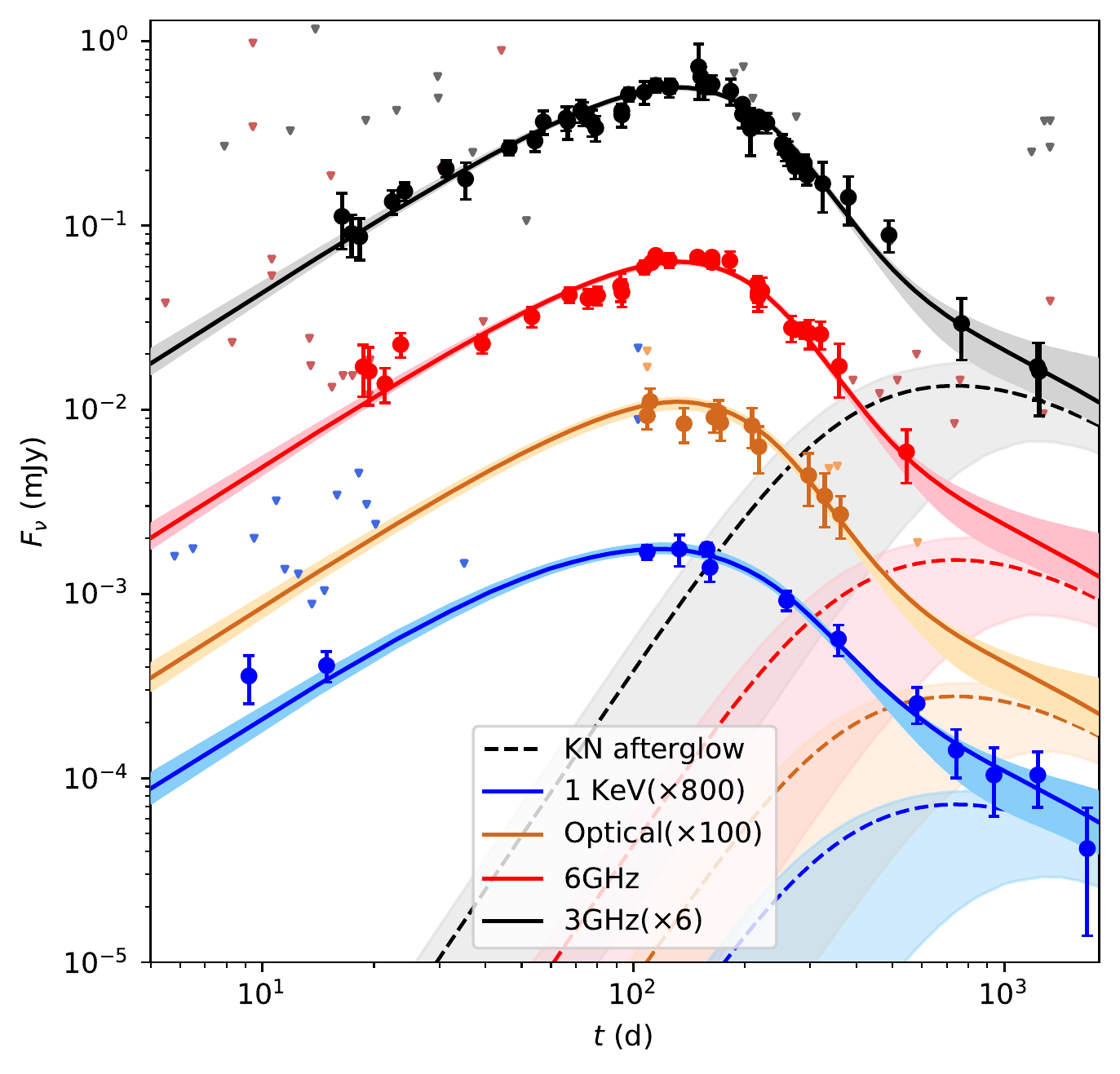}
    \label{subfig:1b}}
	\caption{\small The fit to the afterglow data in the time interval of 5 to 1800 days after GW170817/GRB 170817A. Panels (a) and (b) show the reproduced light curves in two different methods, the boosted-fireball structured-jet model ({\tt JetFit}) and the Gaussian structured jet model ({\tt Afterglowpy}). We denote the frequencies of 1 keV, $5.06 \times 10^{14}$ Hz (optical), 6 GHz, and 3 GHz in black, red, brown, and blue, respectively. Here, each band's flux $F_{\nu}$ is re-scaled for a better view. The solid lines (with maximum likelihood) represent the best fit of the whole afterglow data, including both the GRB and the kilonova afterglow components. 
	The colored regions represent the corresponding $90\%$ credible regions. The dashed lines are the kilonova afterglow components. The light-colored regions represent the $90\%$ credible regions of the estimated kilonova light curves. The observation data points are marked in circles with error bars, and the upper limits are shown in lower triangles.
	}
	\label{fig:LC}
\end{figure}

\begin{figure*}[ht!]
	\centering
	\includegraphics[width=0.9\textwidth]{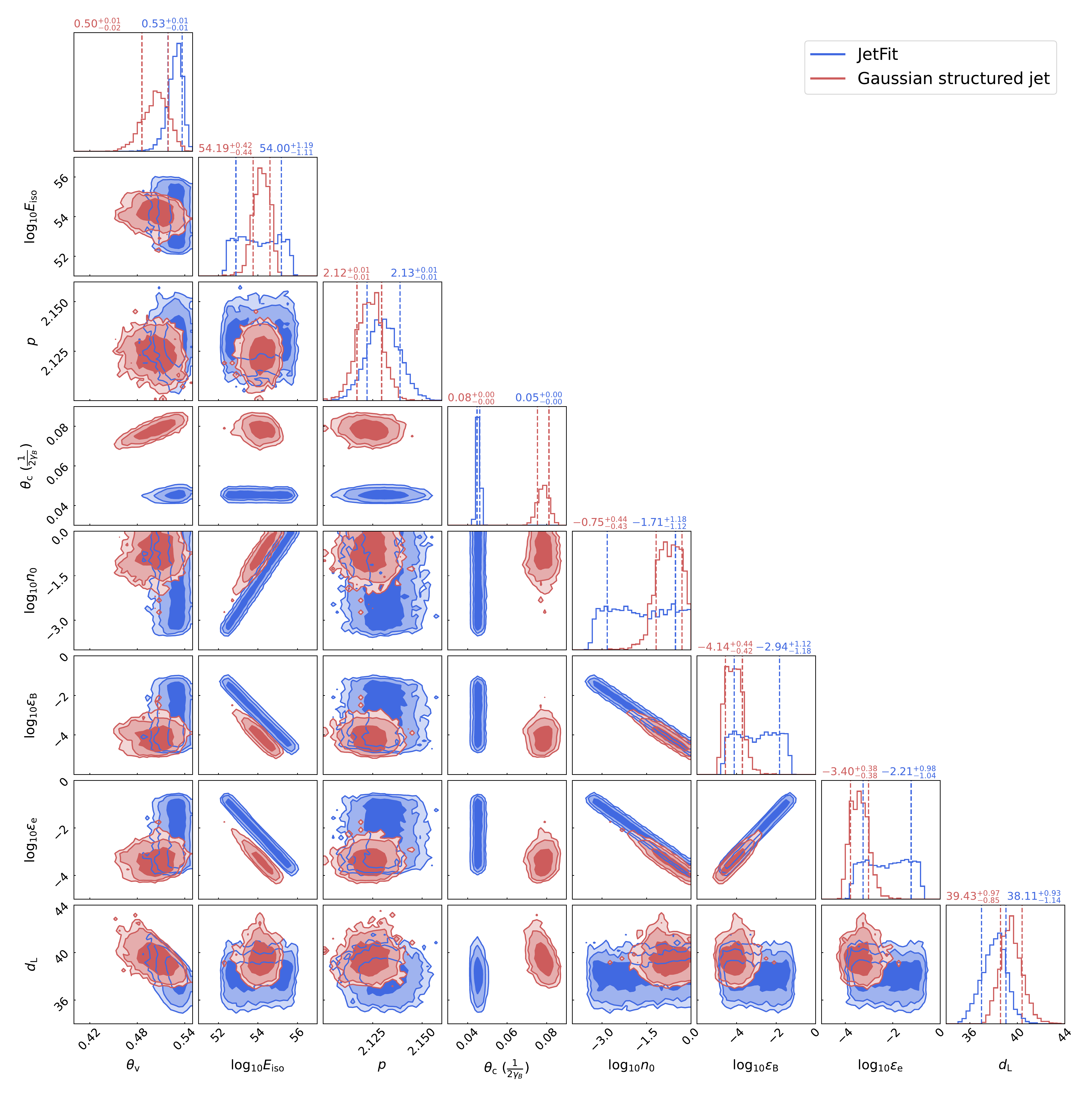}
    \caption{\small Posterior distributions of some key parameters for the best-fit light curves presented in Figure.~\ref{fig:LC}. The results of the {\tt JetFit} and the Gaussian structured models are marked in blue and red, respectively. As the jet opening angel is $\sim 1/\gamma_B$ in the boosted fireball model, we plot the half opening angle $\theta_c$ and ${1}/{2\gamma_B}$ in the same corner for a convenient comparison. In {\tt Jetfit} and Gaussian structured models, the energy of the GRB jet is described as the explosion energy $E_0$ and the isotropic-equivalent energy $E_{\rm iso}$ at $\theta=0$, respectively. Considering $E_{\rm iso} \sim  {2E_0}/{[1-\cos(\theta_0 /2)]}$ \citep{2018ApJ...869...55W}, we convert $E_0$ to $E_{\rm iso}$ for a convenient comparison. The contours are at the $68\%$, $95\%$, $99\%$ credible level. The values are at $68\%$ credible level.}
	\label{fig:posterior}
\end{figure*}

We have systematically investigated the factors that might influence the determination of viewing angle, including the approaches ({\tt Afterglowpy} and {\tt JetFit}), the structured jet models (Gaussian, Power-Law, and Top-Hat structures), and the data sets (200-days/entire afterglow data and VLBI's constraint). Figure.~\ref{fig:LC} presents our optimal fitting with two different models. The posterior results shown in Figure~\ref{fig:posterior} indicate that the superluminal motion of the jet gives a combined limitation between the viewing angle and the luminosity distance. In general, the inclusion of the late time (i.e., $\geq 200$ days) afterglow data yields a larger viewing angle with a lower uncertainty, as anticipated (one exception is the superluminal motion constrains the viewing angle obtained in the {\tt JetFit} model of all the afterglow data to be consistent with that of the first 200 days).
Without the sideways lateral expansion, the logarithm of Bayes evidence (${\rm ln}Z$) is $10$ less than the Gaussian structured jet with expansion, suggesting a much poorer fit. For {\tt JetFit} model using the entire data set, the $\theta_{\rm v}$ is constrained to $0.53^{+0.01}_{-0.01}$ rad (at the $68.3\%$ credible level; other parameters of the afterglow model are presented in Table~\ref{tb:Ag_par} and Figure~\ref{fig:posterior}, which is consistent to the results obtained by \citet{2021MNRAS.502.1843N}.) Using the same boosted fireball model and similar data set, \citet{2022ApJ...927L..17H} instead found $\theta_{\rm v}=0.44^{+0.01}_{-0.01}$ rad. Such a difference is mainly due to the fixed $n_{0,0}$, $\gamma_B$, and $\epsilon_e$ in the afterglow modeling of \citet{2022ApJ...927L..17H}.
With {\tt Afterglowpy}, the Gaussian structured jet model (the posterior results are shown in Figure~\ref{fig:posterior}) gives a $\theta_{\rm v}=0.51^{+0.01}_{-0.02}$ rad. For the {\tt JetFit} and the {\tt Afterglowpy} Gaussian structured jet models, the ${\rm ln}Z$ are 594 and 562, respectively. In comparison to the {\tt Afterglowpy} Gaussian jet model, the Power-Law and Top-Hat models can not well fit the observations, and the results are ${\rm ln}Z = 546$ ($\theta_{\rm v}=0.46^{+0.01}_{-0.01}$ rad) and 426 ($\theta_{\rm v}=0.50^{+0.03}_{-0.04}$ rad), respectively. All of the posterior distributions of these scenarios are presented in Appendix~\ref{sec:6}. Therefore, we only display the best-fit light curves for the {\tt JetFit} and Gaussian structured jet models in Figure~\ref{fig:LC}. We would like to also comment on the role of the kilonova afterglow component. Though the inclusion of this component can improve the goodness of the fit, its existence can not be convincingly established in the {\tt JetFit} model, for which the enhancement of $\ln Z$ is just $\approx 4$. In the {\tt Afterglowpy} Gaussian jet model, the kilonova afterglow is more prominent and dominates the emission at $t\geq 700$ days. The corresponding posterior distribution is presented in Appendix~\ref{sec:6} (The estimated parameters of kilonova afterglow do not show well convergence if we use {\tt JetFit} model for GRB afterglow). Motivated by its highest $\ln Z$ value, in this work we adopt the {\tt JetFit} model as our fiducial approach. Nevertheless, the rather similar $\theta_{\rm v}$ inferred in the above diverse approaches/models do consistently favor a large viewing angle of $\approx 0.5$ rad. Another thing that should be mentioned is that the explosion energy we obtained is larger than that estimated in several early works \citep[e.g.,][]{2019ApJ...870L..15L,2019MNRAS.485.2155L,2019MNRAS.489.1919T,2020ApJ...896..166R}, but is close to some recent 
evaluations  \citep[e.g.,][]{2021MNRAS.502.1843N,2022ApJ...927L..17H}. Such high energy supports the claim that GRB 170817A would be one of the brightest short events detected so far \citep{DuanKK2019}. The inferred large $\theta_{\rm v}$ points towards a high GRB/GW association rate and hence a promising multi-messenger detection prospect of the double neutron star mergers in the future \citep{2018ApJ...857..128J}.
Besides, we do not find evidence for the sizeable deviation of the model prediction from the data, which suggests that there is no sign of a continual energy injection from the central engine. Hence, it is in favor of a black hole rather than a neutron star central engine for GRB 170817A \cite[see also][for an independent argument]{2022arXiv220713613H}.

\begin{figure*}[ht!]
	\centering
	\includegraphics[width=0.8\textwidth]{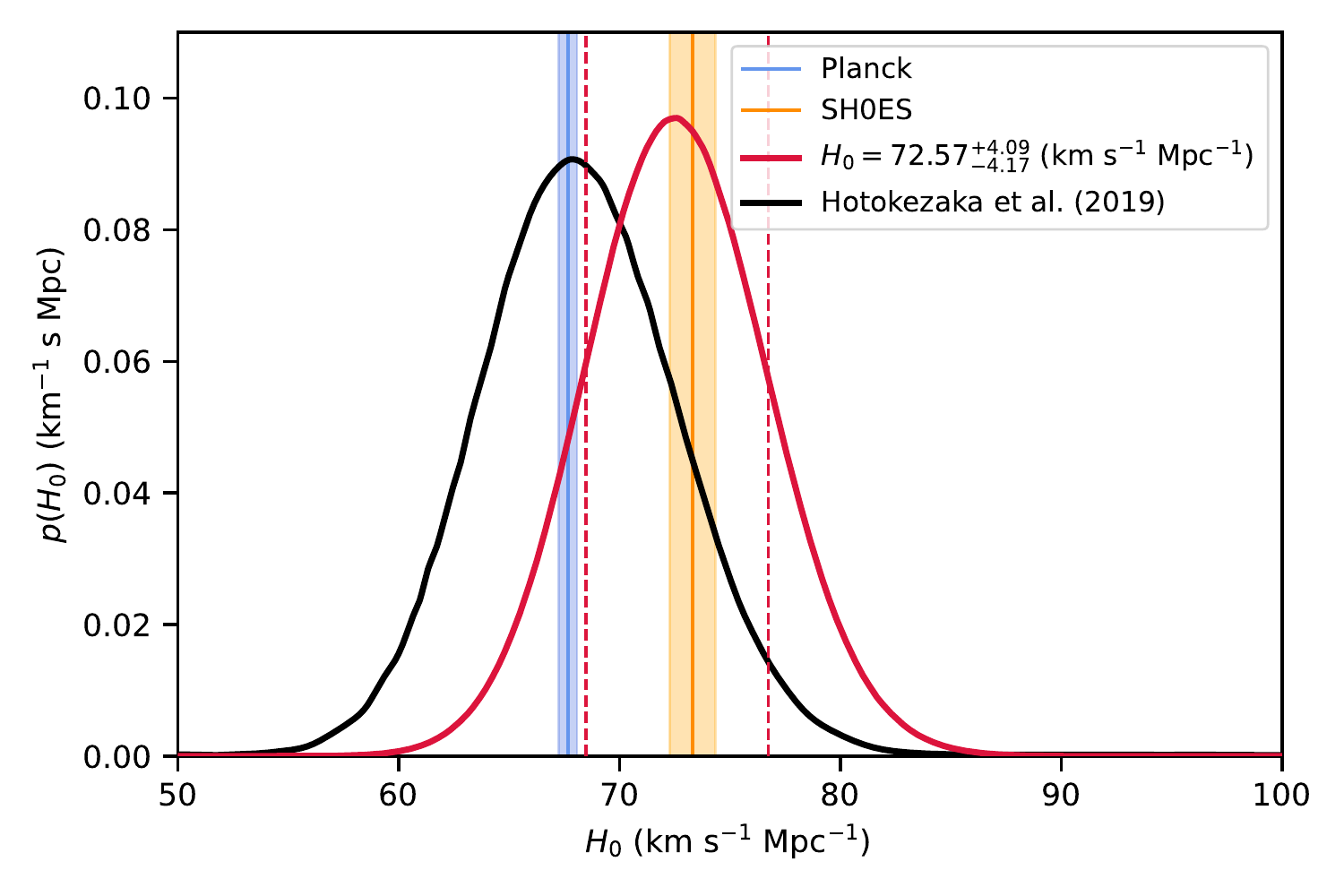}
    \caption{\small The Hubble constant inferred with the data of GW170817/GRB 170817A. The red solid line represents the $H_0$ estimated in {\tt JetFit} model ($H_0=72.57^{+4.09}_{-4.17}\, \rm km\, s^{-1}\, Mpc^{-1}$) and the black solid line represents the result from \citet{2019NatAs...3..940H}, while the blue and yellow regions represent the measurement results of Planck ($67.66 \pm 0.42 {\rm ~km\,s^{-1}\,Mpc^{-1}}$) and SH0ES ($73.30 \pm 1.04 {\rm ~km\,s^{-1}\,Mpc^{-1}}$), respectively.}
	\label{fig:H0}
\end{figure*}

Then, we use the posterior distribution of the viewing angle $\theta_{\rm v}=0.53^{+0.01}_{-0.01} \rm rad$ obtained in {\tt JetFit} model ($\theta_{\rm v}=0.51^{+0.01}_{-0.02}\rm rad$ for Gaussian jet model) as the prior of inclination angle in GW data analysis. The full Bayesian inference on GW170817 limits $d_{\rm L}$ to $40.67^{+1.11}_{-1.03}\,\rm Mpc$ ($d_{\rm L}=41.09^{+1.19}_{-1.05}\,\rm Mpc$ when using Gaussian structured jet) at the $68.3\%$ confidence level. The estimations of other GW parameters are shown in Table.~\ref{tb:GW_par}. Because of the breaking of the degeneracy between $\iota$ and $d_{\rm L}$, GW analysis gives about $5.6\%$ uncertainty on the Hubble constant and limits $H_0$ close to the SH0ES result, i.e., $H_0=72.57^{+4.09}_{-4.17}\, \rm km\, s^{-1}\, Mpc^{-1}$ ($H_0=71.80^{+4.15}_{-4.07}\, \rm km\, s^{-1}\, Mpc^{-1}$ when using Gaussian jet model) at the $68.3\%$ credible level (see Figure~\ref{fig:H0}). If inheriting both the estimated $d_{\rm L}$ and $\theta_{\rm v}$ from afterglow fitting as the prior of GW analysis, we would have  $H_0=75.52^{+4.09}_{-4.04}\, \rm km\, s^{-1}\, Mpc^{-1}$ ($H_0=73.53^{+4.00}_{-3.96}\, \rm km\, s^{-1}\, Mpc^{-1}$) for {\tt JetFit} (Gaussian) model, with which the difference from the Planck $H_0$ measurement is more distinct.
One thing that should be mentioned is that if we do not take into account the constraint of viewing angle, i.e., $0.25<\theta_{\rm v} \bigl(\frac{d_{\rm L}}{41 \rm Mpc} \bigr)<0.5$ rad \citep{2018Natur.554..207M, 2019NatAs...3..940H}, in the afterglow modeling, $\theta_{\rm v}$ can still be well constrained but prefers a larger value ($\sim 0.6 \rm \, rad$ in {\tt JetFit} model) and hence a larger $d_{\rm L}$. Correspondingly, in comparison to the results considering the VLBI's constraint, we would yield a larger result $H_0=74.36^{+4.44}_{-4.32}\, \rm km\, s^{-1}\, Mpc^{-1}$, and the difference from the Planck $H_0$ measurement is more distinct, also.

\begin{table*}[ht!]
\begin{ruledtabular}
\centering
\caption{Prior distributions and posterior results of the parameters for GW170817}
\label{tb:GW_par}
\begin{tabular}{lccc}
Names                   &Parameters                   &Priors of parameter inference         &Posterior results\textsuperscript{b}\\ \hline    
Chirp mass              &$\mathcal{M}/M_{\odot}$      &Uniform(0.4, 4.4)     &$1.1976^{+0.0001}_{-0.0001}$\\
Mass ratio              &$q$                          &Uniform(0.125, 1.0)   &$0.76^{+0.16}_{-0.17}$\\
Aligned spin  &$\chi_{1,2}$              &AlignedSpin\textsuperscript{a}  &$0.01^{+0.12}_{-0.09}~\&~0.02^{+0.16}_{-0.14}$\\
Polarization of GW      &$\psi$                       &Uniform(0, 2$\pi$)   &$1.60^{+1.04}_{-1.16}$\\	
Coalescence time        &$t_{\rm c}/\rm s$            &1187008882.42 &-\\
Coalescence phase       &$\phi_{\rm c}$               &Uniform(0, 2$\pi$)   &Marginalized         \\
Right ascension         &$\alpha$                     &3.44616      &-\\
Declination             &$\delta$                     &-0.408084    &-  \\
Tidal deformability     &$\Lambda_{1,2}$              &Uniform(0,5000) &$204^{+342}_{-147}~\&~590^{+752}_{-344}$\\
Inclination angle       &$\iota/\rm rad$                      &Constrained by afterglow model analysis &$2.61^{+0.01}_{-0.01}$\\
Luminosity distance     &$d_{\rm L}/\rm Mpc$          &Gaussian($\mu=40.7, \sigma=2.36$) &$40.67^{+1.11}_{-1.03}$\\
\end{tabular}
\begin{tablenotes}
  \item[a] \textsuperscript{a} The spin component projected to the orbit angular momentum follows the distribution described in Equation (A7) of \citet{2018arXiv180510457L} with $\chi_{\rm max}$ = 0.89, and the spin's tilt angle is taken to be aligned.
  \item[b] \textsuperscript{b} The posterior results are at the $68.3\%$ credible level.
\end{tablenotes}
\end{ruledtabular}
\end{table*}

\section{Conclusion}\label{sec:5}
For the GW event accompanying an EM counterpart, $H_0$ can be estimated by the standard siren method. This method utilizes redshift information from a confirmed host galaxy and the posterior distribution of luminosity distance from GW data analysis to obtain the $H_0$ value following Hubble's Law. The bright BNS merger event GW170817 has been identified in NGC 4993, so the Hubble flow velocity can be measured. After fitting the multi-wavelength light curves with the model developed by \citet{2018ApJ...869...55W}, the viewing angle $\theta_{\rm v}$ is constrained to $0.53^{+0.01}_{-0.01} \, \rm rad$ ($\theta_{\rm v}=0.51^{+0.01}_{-0.02}$), which partially breaks the degeneracy between $\iota$ and $d_{\rm L}$ and improves the accuracy of $d_{\rm L}$ estimation in GW data analysis. Therefore, we finally obtain the estimation of the Hubble constant as $H_0=72.57^{+4.09}_{-4.17}\, \rm km\, s^{-1}\, Mpc^{-1}$ ($H_0=71.80^{+4.15}_{-4.07}\, \rm km\, s^{-1}\, Mpc^{-1}$) from GW170817/GRB170817, which is more consistent with the SH0ES result rather than the CMB result. However, the uncertainty is still too large to confirm the Hubble tension. In our modeling, the possible kilonova afterglow component has been taken into account. In the {\tt JetFit} model, the contribution of such a new component is not significant. In the {\tt Afterglowpy} Gaussian structured jet model, the kilonova afterglow is more prominent and dominates the observed flux at $t\geq 700$ days. This distinction is most likely arisen from the different treatments on the sideways expansion. The {\tt Afterglowpy} assumes the expansion spreads at sound speeds \citep{2020ApJ...896..166R}, while the {\tt JetFit} model has a slower speed based on relativistic hydrodynamical jet simulations. 
Therefore, in the {\tt JetFit} scenario, the deceleration of the GRB ejecta will be less prominent than the case of {\tt Afterglowpy} \citep{2003ApJ...591.1075K,2012ApJ...749...44V,2018ApJ...869...55W}. Consequently, the GRB afterglow emission decline is shallower for {\tt JetFit} and the contribution of the kilonova afterglow component is suppressed. Therefore, more data are needed to convincingly establish the presence of a kilonova afterglow.

With the improvement of multi-band telescopes and the upgrade of GW interferometers, it is expected to detect more neutron star mergers with multi-messengers. \citet{2018Natur.562..545C} have predicted that $H_0$ measurement will reach two percent precision within five years based on the standard siren method. And the combination of posterior distribution of $H_0$ estimation can reduce the error of $H_0$ to $\sigma_{H_0}/\sqrt{N}$ with $N$ bright GW events, where $\sigma_{H_0}$ is the typical width of the $H_0$ measurement. As shown in \citet{2020MNRAS.493.1633S} and \citet{2022MNRAS.513.4159P}, the number of multi-messenger detection of BNS merger will be close to the single digits during O4/O5 because the large $d_{\rm L}$ restricts the detection of GW signal and the large $\theta_{\rm v}$ restricts the GRB detection for the meanwhile. Since GW170817 is very close to us, its off-axis afterglow emission is still detectable in quite a few years. However, it is not the case for more distant events as expected. 
Fortunately, the afterglow emission would be much brighter for the on-axis events. More importantly, the jet opening angle, as well as its uncertainty, can be reliably estimated, with which both the $d_{\rm L}$ and $H_0$ can be well constrained ($\sim 3\%$ accuracy) even with a single BNS merger \citep{2022PhRvD.106b3011W}. In view of the above facts, we conclude that more precise $H_0$ is expected with the GW standard sirens in the near future, and the Hubble tension will be credibly clarified.

\begin{acknowledgements}
We thank the anonymous referee for very helpful comments and suggestions. Y.Y. Wang thanks S.J. Gao for the help in improving the computational efficiency of the codes and for the valuable suggestions. This work was supported in part by NSFC under Grants No. 11921003, No. 12225305 and No. 12233011. This research has made use of data and software obtained from the Gravitational Wave Open Science Center \url{https://www.gw-openscience.org}, a service of LIGO Laboratory, the LIGO Scientific Collaboration, and the Virgo Collaboration. LIGO is funded by the U.S. National Science Foundation. Virgo is funded by the French Centre National de Recherche Scientifique (CNRS), the Italian Istituto Nazionale della Fisica Nucleare (INFN), and the Dutch Nikhef, with contributions by Polish and Hungarian institutes. 

$Software:$ {\tt Afterglowpy} (\citet{2020ApJ...896..166R}, \url{https://pypi.org/project/afterglowpy/}), {\tt JetFit} (\citet{2018ApJ...869...55W}, \url{https://github.com/NYU-CAL/JetFit}), {\tt Bilby} (\citet{2019ApJS..241...27A}, version 1.0.4, \url{https://git.ligo.org/lscsoft/bilby/}), {\tt Pymultinest} (\citet{2016ascl.soft06005B}, version 2.11, \url{https://pypi.org/project/pymultinest/}), {\tt Dynesty} (\citet{2020MNRAS.493.3132S}, version 1.1, \url{https://dynesty.readthedocs.io/en/latest/}).
\end{acknowledgements}

\appendix
\section{The posterior distributions of the GRB and kilonova afterglow parameters}\label{sec:6}
\begin{figure*}[ht!]
	\centering
	\includegraphics[width=0.8\textwidth]{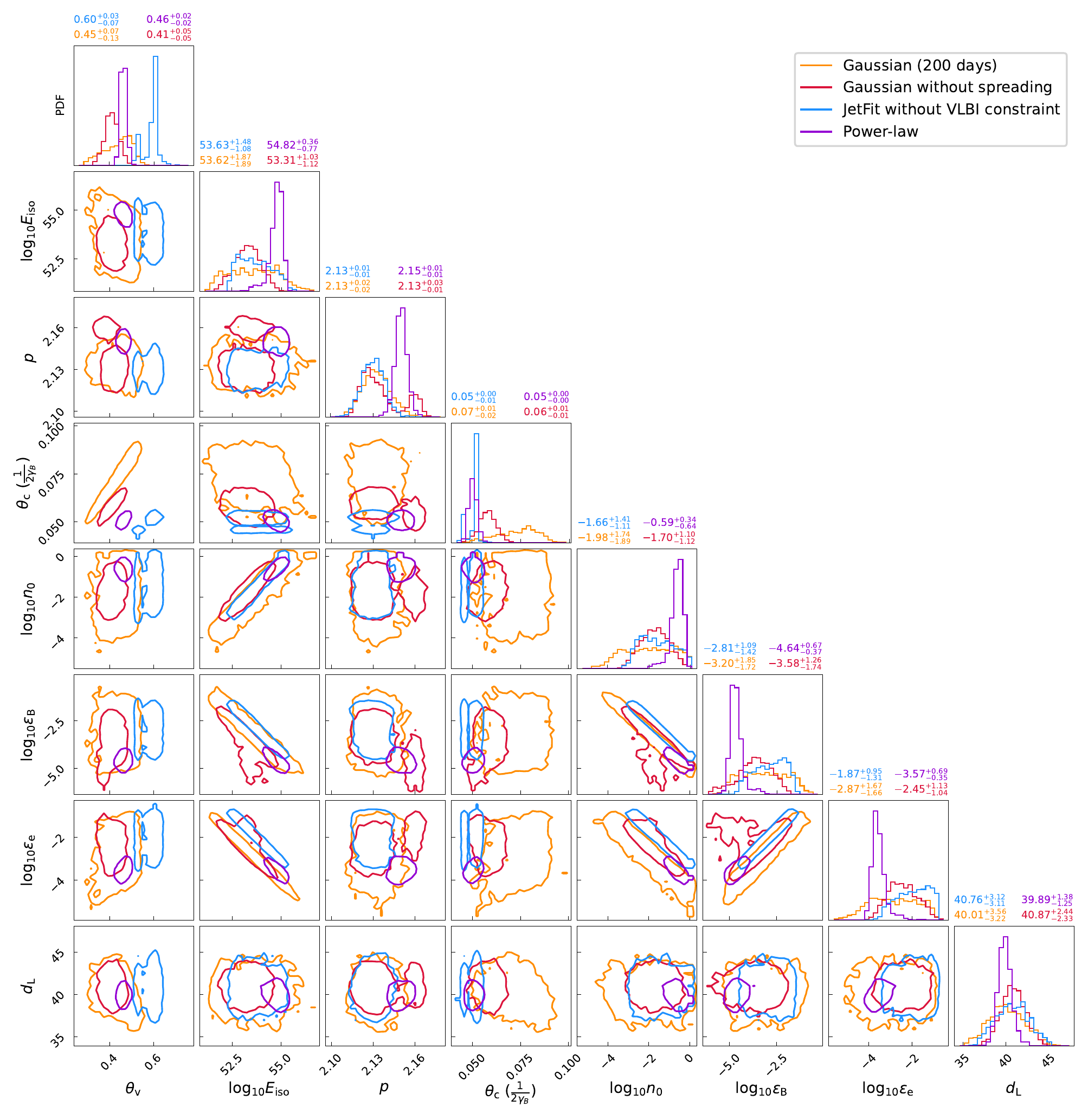}
    \caption{\small The posterior distributions of the parameters of four afterglow modeling. The {\tt JetFit} model without the constraint of the superluminal motion (${\rm ln}Z = 596$), the Gaussian model without lateral spreading (${\rm ln}Z = 553$), and the Power-Law model (${\rm ln}Z = 546$) are represented in blue, red and purple, respectively. The Gaussian model just with the first 200 day observation data (${\rm ln}Z = 301$) is also shown in orange for comparison. In these scenarios, the kilonova afterglow component has been taken into account. 
     All of the ranges are at the $90\%$ credible level.}
	\label{fig:app}
\end{figure*}

\begin{figure*}[ht!]
	\centering
	\includegraphics[width=0.8\textwidth]{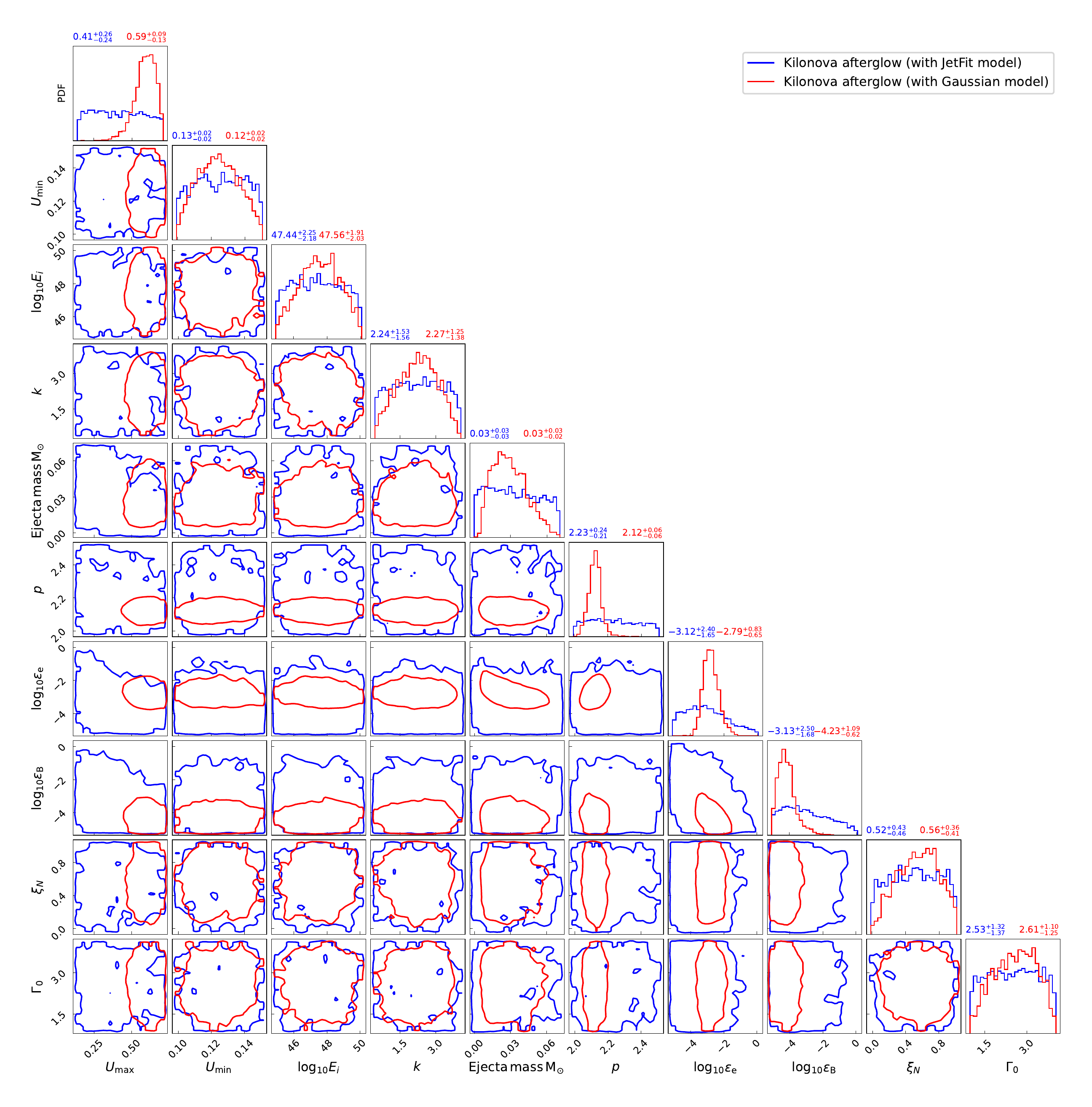}
    \caption{\small The posterior distributions of some parameters of kilonova afterglow modeling (see Figure.~\ref{fig:LC} for the reproduced light curves). Note that the posterior distributions of the circumburst density $n_0$ and the luminosity distance $d_{\rm L}$ have already been presented in Figure.~\ref{fig:posterior}. The ranges are at the $90\%$ credible level.}
	\label{fig:app2}
\end{figure*}

Here, we present some posterior distributions of the GRB and kilonova afterglow parameters mentioned in Section.~\ref{sec:4}. As an extra supplementary for previous discussions, Figure.~\ref{fig:app} shows four scenarios, including the {\tt JetFit} model without the constraint of the superluminal motion, the Gaussian structured jet model without the lateral spreading, the Gaussian structure jet model with fitting data during the first 200 days only, and the Power-Law structured jet. The first scenario yields a high $\ln Z$ (the same as the {\tt JetFit} model with the constraint of the superluminal motion) but would predict an even higher $H_0$ because of the suggested $\theta_{\rm v}\sim 0.6$ rad. The last three scenarios have significantly lower $\ln Z$ (represented in Figure.~\ref{fig:app}) because of the poorer fits to the data. 
In Figure.~\ref{fig:app2}, we plot the posterior distributions of the parameters of kilonova afterglow modeling displayed in Figure.~\ref{fig:LC}. The spherical cocoon model, which describes the evolution of the kilonova afterglow approximately, specifies that the energy-velocity distribution follows a Power-Law distribution $E(u)=E_0 ({u}/{u_{\rm max}})^{-k}$, where $u$ is the dimensionless 4-velocity and within $(u_{\rm min}, u_{\rm max})$. In this framework, the shock driven by the kilonova blast wave is refreshed by the coasting of the slow material when it decelerates.

\clearpage
\bibliography{ref}
\bibliographystyle{aasjournal}
\end{document}